\pdfoutput=1
\documentclass[aps,prb,twocolumn,groupedaddress,showpacs]{revtex4-1}
\usepackage{amsmath,amssymb,latexsym,epic,eepic,epsfig,graphicx,psfrag}
\usepackage[latin1]{inputenc}

\usepackage{revsymb4-1}
\usepackage{bm}
\usepackage{comment}

\begin{document}

%\preprint{}

\title{Semi-Empirical Model for Nano-Scale Device Simulations}


\author{Kurt Stokbro}
\homepage{http://www.quantumwise.com}
\email[]{kurt.stokbro@quantumwise.com}
\author{Dan Erik Petersen}
\author{S{\o}ren Smidstrup}
\author{Anders Blom}
\author{Mads Ipsen}
\affiliation{QuantumWise A/S,\\
 N{\o}rre S{\o}gade 27A, 1.~th,
  DK-1370 Copenhagen K, Denmark}
\author{Kristen Kaasbjerg}
\affiliation{Center for Atomic-scale Materials Design (CAMd),\\
  Department of Physics, Technical University of Denmark,\\
  DK-2800 Kgs.\ Lyngby, Denmark}

%\thanks{}

\date{\today}

\begin{abstract}
We present a new semi-empirical model for calculating
electron transport in atomic-scale devices. The model is an
extension of the Extended H{\"u}ckel method with a self-consistent Hartree
potential. This potential models the effect of an external bias and
corresponding charge re-arrangements in the device. It is also
possible to include the effect of external gate potentials and
continuum dielectric regions in the device. 
The model is used to study the
electron transport through an organic molecule between gold surfaces,
and it is demonstrated that the results are in closer agreement with
experiments than \textit{ab initio} approaches provide. In another example, we study the
transition from tunneling to thermionic emission in a transistor 
structure based on graphene nanoribbons.
\end{abstract}

%\keywords{}

\maketitle

\section{\label{sec:intro}Introduction}

As the minimum feature sizes of electronic devices are approaching the
atomic scale, it becomes increasingly important to include the effects of single
atoms in device simulations. In recent years, there have been several
developments of atomic-scale electron transport simulation models based on the
Non-Equilibrium Green's Function (NEGF) formalism\cite{Haug1996}. The approaches can 
roughly be divided into two catagories: \textit{ab initio} approaches, where
the electronic structure of the system is calculated from first principles, typically with
Density Functional Theory (DFT)\cite{Lang1995,Xue2002,Brandbyge-2002-PRB,Taylor2001},
and semi-empirical approaches, where the electronic structure is calculated
using a model with adjustable parameters fitted to experiments or
first-principles calculations. Examples of semi-empirical transport models are
methods based on Slater-Koster tight-binding
parameters\cite{Carlo2002,Pecchia-2004-RPP} and Extended H{\"u}ckel
parameters\cite{MaJo97,CoCeSa99,Cerda2000,EmKi01,Zahid2005,KienleI2006,KienleII2006}.

The \textit{ab initio} models have the advantage of predictive power, and can
often give reasonable results for systems where there is no prior experimental
data. However, the use of the Kohn-Sham one-particle states as quasiparticles is
questionable, and it is well known that for many systems the energies of the
unoccupied levels are rather poorly described within DFT. Furthermore, solving
the Kohn-Sham equations can be computationally demanding, and solving for device
structures with thousands of atoms is only feasible on large parallel
computers.

The semi-empirical models have less predictive power, but when used within their
application domain they can give very accurate results. The models may also be
fitted to experimental data, and can thus in some cases give more accurate results
than DFT-based methods. However, the main advantage of the semi-empirical
methods are their lower computational cost.

In this paper we will present the formalism behind a new semi-empirical
transport model based on the Extended H{\"u}ckel (EH) method. The model can be viewed as an
extension to the work by Zahid \textit{et al.}\cite{Zahid2005}, with the main
difference being the treatment of the electrostatic interactions. Zahid \textit{et al.}
only describe part of the electrostatic interactions in the device; most
importantly, they use the Fermi level of the electrodes as a fitting parameter and
do not account for the charge transfer from the electrodes to the device. In the
current work, the Fermi level of the electrodes is determined self-consistently
by using the methodology introduced by Brandbyge \textit{et
 al.}\cite{Brandbyge-2002-PRB} In this way, we include the charge transfer from
the electrodes to the device region and describe all electrostatic interactions
self-consistently. This is accomplished by defining a real-space electron
density and numerically solving for the Hartree potential on a real-space grid.
Through a multi-grid Poisson solver, we include the self-consistent field from an
applied bias, and allow for including continuum dielectric regions and
electrostatic gates within the scattering region.

The organization of the paper is the following: In section~\ref{sec:theory} we
introduce the self-consistent Extended H{\"u}ckel (EH-SCF) model, and in
section~\ref{sec:method} we present the formalism for modelling nano-scale
devices. In section~\ref{sec:tourwire} we apply the model to a
molecular device, and in section~\ref{sec:graphene} we consider a
graphene nano-transistor where an electrostatic gate is controlling the electron
transport in the device. Finally, in section~\ref{sec:summary}, we conclude the paper.

\section{\label{sec:theory}The Self-Consistent Extended-H{\"u}ckel method}

In this section we describe the EH-SCF framework. In Extended H{\"u}ckel theory, the
electronic structure of the system is expanded in a basis set of local atomic
orbitals (LCAOs)
\begin{equation}
  \phi_{nlm}({\bf r}) = R_{nl}(r) Y_{lm}(\hat{r}),
\end{equation}
where $Y_{lm} $ is a spherical harmonic and $R_{nl}$ is a superposition of
Slater orbitals
\begin{equation}
  R_{nl}(r) = \frac{r^{n-1-l}}{(2n)!}  \left[ C_1 (2 \eta_1)^{2n+1} e^{-\eta_1 \, r}+C_2 (2 \eta_2)^{2n+1} e^{-\eta_2 \, r} \right]. 
\end{equation}
The LCAOs are described by the adjustable parameters $\eta_1$, $\eta_2$, $C_1$, and
$C_2 $, and these parameters must be defined for the valence orbitals of each
element.

The central object in EH theory is the overlap matrix,
\begin{equation}
S_{ij}  = \left\{ 
\begin{array}{l l}
 \delta_{ ij}    & \quad \mbox{if ${\bf R}_i={\bf R}_j$} \\
  \int_V \phi_i({\bf r} -{\bf R}_i) \phi_j({\bf r}- {\bf R}_j) \; \mathrm{d}{\bf r} & \quad \mbox{if ${\bf R}_i \neq {\bf R}_j$}\\
  \end{array}
\right.
\end{equation} 
where $i$ is a composite index for $nlm$ and ${\bf R}_i$ is the position of the
center of orbital $i$.
 
From the overlap matrix, the one-electron Hamiltonian is defined by
\begin{equation}
  H_{ij}= \left\{ 
\begin{array}{l l}
 E_i + \delta V_H({\bf R}_i)  & \quad \mbox{if $ i=j$} \\
 \frac{1}{4} (\beta_i+\beta_j) (E_i+E_j) S_{ij} \, +  \, \frac{1}{2} \left(\delta V_H({\bf R}_i)+\delta V_H({\bf R}_j)\right)S_{ij} &  \quad \mbox{if $ i \neq j$}\\
  \end{array}
\right.
\label{eq:EH_hamiltonian}
\end{equation} 
where $E_i $ is an orbital energy, and $\beta_i$ is an adjustable parameter
(often chosen to be 1.75). $\delta V_H({\bf R}_i) $ is the Hartree potential
corresponding to the induced electron density on the atoms, i.e.\ the change in
electron density compared to a superposition of neutral atomic-like densities.
This term must be determined self-consistently, and is not included in 
standard EH models\cite{Hoffmann1978}. In the following section we describe how this term is
calculated.

\subsection{Solving the Poisson Equation to Obtain the Hartree Potential}
      
To calculate the induced Hartree potential we need to determine the spatial
distribution of the electron density. To this end, we introduce the Mulliken
population of atom number $\mu $
\begin{equation}  
  m_\mu =  \sum_{i \in \mu} \sum_j D_{ij} S_{ij},
\end{equation}  
where $D_{ij}$ is the density matrix. The total number of electrons can now be
written as a sum of atomic contributions, $N=\sum_\mu m_\mu $.

We will represent the spatial distribution of each atomic contribution by a
Gaussian function, and use the following approximation for the spatial
distribution of the induced electron density:
\begin{equation}  
  \delta n({\bf r})=\sum_\mu \delta m_\mu \sqrt{
    \frac{\alpha_{\mu}}{\pi}} e^{-\alpha_\mu |{\bf r}-{\bf R}_\mu|^2},
\label{eq:realspacedens}
\end{equation}  
where the weight $\delta m_\mu = m_\mu - Z_\mu $ of each Gaussian is the
excess charge of atom $\mu$ as obtained from the Mulliken population
$m_\mu $ and the ion valence charge $Z_\mu $. Subsequently, the Hartree
potential is calculated from the Poisson equation
\begin{equation}
  \label{eq:poisson}
  - \nabla \cdot \left[ \epsilon(\mathbf{r})\, \nabla \delta V_{H}(\mathbf{r}) \right] = \delta n(\mathbf{r}),
\end{equation}
which is solved with the appropriate boundary conditions on the leads and gate electrodes
imposed by the applied voltages. Here, $\epsilon(\mathbf{r})$ is the spatially dependent
dielectric constant of the device constituents, and allows for the inclusion
of dielectric screening regions.

To see the significance of the width $\alpha_\mu $ of the Gaussian orbital, let us 
calculate the electrostatic potential from a single Gaussian electron
density at position ${\bf R}_\mu$. The result is
\begin{equation}  
  \delta V_{H}({\bf r}) =   e (m_\mu-Z_\mu) \frac{\text{Erf}(\sqrt{\alpha_\mu} | {\bf r}-{\bf R}_\mu|) }{ | {\bf r}-{\bf R}_\mu| },
\end{equation}  
and from this equation we see that the on-site value of the Hartree potential is
$\delta V_{H}({\bf R}_\mu)= (m_\mu-Z_\mu) \gamma_\mu $, where the parameter
\begin{equation}
\gamma_\mu= 2 e \sqrt{ \frac{\alpha_{\mu}}{\pi}}
\label{eq:alpha}
\end{equation}
is the on-site Hartree shift. The parameter $\gamma_\mu$ is a well-known quantity
in CNDO theory\cite{Pople1966,Murell1972}, and values of $\gamma_\mu$ are listed
for many elements in the periodic table. Thus, we fix the value of $\gamma_\mu$
using CNDO theory, and then use Eq.~(\ref{eq:alpha})
to calculate the value of $\alpha_\mu$ for each element.

\section{\label{sec:method} EH-SCF Method for a Nano-Scale Device}

\begin{figure}[tbp]
\begin{center}
\includegraphics[width=\linewidth]{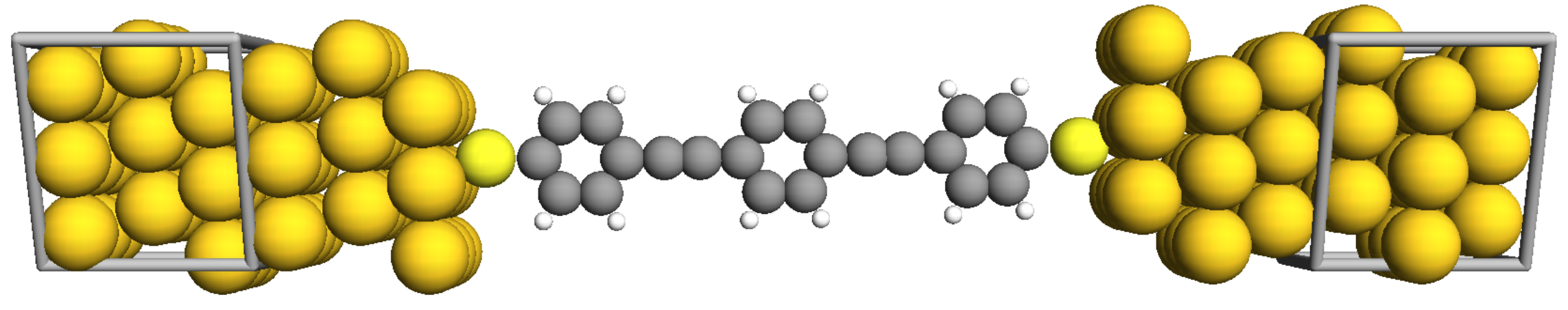}
\end{center}
\caption{(Color online) Geometry of a nano-device consisting of a
  dithiol-triethynylene-phenylene molecule attached to two (3x3) (111) gold
  electrodes. The left and right electrode regions are illustrated with wire
  boxes, and the properties of these regions are obtained from a calculation
  with periodic boundary conditions in all directions. The region between the
  two electrodes is the central device region, which is described with open
  boundary conditions in the transport direction, and periodic boundary
  conditions in the directions perpendicular to the transport direction.  }
\label{fig:sym-device2}
\end{figure}

Fig.~\ref{fig:sym-device2} illustrates the setup of a molecular
device system. The system consists of three
regions: the central region, and the left and right electrode regions. The
central region includes the active parts of the device and sufficient
parts of the contacts, such that the properties of the electrode
regions can be described as bulk materials. For metallic contacts, this
will typically be achieved by extending the central region 5--10 \AA\ into the
contacts.

The calculation of the electron transport properties of the system
is divided into two parts. The first part is a self-consistent calculation for the
electrodes, with periodic boundary conditions in the transport direction. In the
directions perpendicular to the transport direction, we apply the same
boundary conditions for the two electrodes and the central region, and
these boundary conditions are described below. 

In the second part of the calculation, the electrodes define the boundary
conditions for a self-consistent open boundary calculation of the properties of the central
region. The main steps in the open boundary calculation is the determination of
the density matrix, the evaluation of the real-space density, and, finally, the
calculation of the Hartree potential. These steps 
will be described in more detail in the following section.

\subsection{Calculating the Self-Consistent Density Matrix of the Central Region}
\label{sec:dm}
In this section we will describe the calculation of the density matrix of the
central region. We assume that the self-consistent properties of the left and right electrodes
have already been obtained, and thus we also know their
respective Fermi levels, $\varepsilon_L^F$ and $\varepsilon_R^F$. We allow for an
external bias $V_b$ to be applied between the two electrodes, and define the
left and right chemical potentials $\mu_L = \varepsilon_L^F-e V_b$ and $\mu_R =
\varepsilon_L^F$. The applied bias thus shifts all energies in the left
electrode, and a positive bias gives rise to an electrical current from left to
right.

The density matrix for this non-equilibrium system, with two different chemical potentials,
is found by filling up the left and right originating states according to
their respective chemical potentials\cite{ToHoSu00,Brandbyge-2002-PRB},
\begin{equation}
\hat{ D}= \int_{-\infty}^{\infty} \left[
\hat{\rho}^L(\varepsilon)n_F(\varepsilon-\mu_L)+ 
\hat{\rho}^R(\varepsilon)n_F(\varepsilon-\mu_R) \right]\, d\varepsilon ,
\label{eq:Dspec}
\end{equation}
where $\hat{\rho}^L$ ($\hat{\rho}^R$) is the contribution to the spectral density of states from
scattering states originating in the left (right) reservoir.

The calculation of the spectral densities is performed using NEGF 
theory, and we write the partial spectral densities as
\begin{equation}
\hat{\rho}^{L,R} (\varepsilon)  = \frac{1}{2 \pi}  
 \hat{G}(\varepsilon) \, \hat{\Gamma}^{L,R}(\varepsilon)
 \,  \hat{G}^{\dagger }(\varepsilon), \label{eq:rhol} \\
\end{equation}
where $\hat{G}$ is the retarded Green's function of the central region, and the
broadening function $\hat{\Gamma} = i [\hat{\Sigma}-\hat{\Sigma}^{\dagger}]$
is given by the self energies $\hat{\Sigma}^L$ and $\hat{\Sigma}^R$, which arise due to the
coupling of the central region with the semi-infinite left and right electrodes, respectively.

Further details of the NEGF formalism can be found
in Refs.~\onlinecite{Haug1996},\onlinecite{Brandbyge-2002-PRB}. Here we just note that to
improve the numerical efficiency, the integral in Eq.~(\ref{eq:Dspec}) is divided
into an equilibrium and non-equilibrium part. The equilibrium part is calculated
on a complex contour far from the real-axis poles of the Green's function,
and only the non-equilibrium part is performed along the real axis. The equilibrium
and non-equilibrium parts are then joined using the double-contour technique
introduced by Brandbyge \textit{et al.}\cite{Brandbyge-2002-PRB}

From the density matrix we may now evaluate the real-space density in the
central region using Eq.~(\ref{eq:realspacedens}). It is important to note that
near the left and right faces of the central region there will be contributions
from the electrode regions, and this ``spill in'' must be properly accounted
for.

Once the real-space density is known, the Hartree potential is calculated by
solving the Poisson equation in Eq.~\eqref{eq:poisson} using a real-space
multi-grid method. On the left and right faces of the central region the Hartree
potential is fixed by the electrode Hartree potentials, appropriately shifted
according to the applied bias. In the directions perpendicular to the transport
directions, we apply the appropriate boundary conditions, fixed or periodic,
as demanded by e.g.\ the presence of gate electrodes.

The so-obtained Hartree potential defines a new Hamiltonian, via Eq.~\ref{eq:EH_hamiltonian}, 
and the steps in section~\ref{sec:dm} must be repeated until a self-consistent
solution is obtained. 

\subsection{Transmission and Current}

Once the self-consistent one-electron Hamiltonian has been obtained, we can
finally evaluate the transmission coefficients~\cite{Haug1996,Datta1995}
\begin{equation}
T(\varepsilon)= {\rm{Tr}}
 [\hat{ \Gamma}_L(\varepsilon)
 \hat{ G}^\dagger(\epsilon)
\hat{ \Gamma}_R(\varepsilon)
 \hat{ G}(\varepsilon) ] \,
\label{eq:orthcond}
\end{equation}
and the current
\begin{equation}
I= \frac{2 e}{h}\int_{-\infty}^{\infty}  T(\varepsilon)[ n_F(\varepsilon-\mu_L)-  n_F(\varepsilon-\mu_R)]\, d \varepsilon.
\label{eq:I2} 
\end{equation}

In the following sections, we apply this formalism to the
calculation of the electrical properties of a molecule between gold
electrodes, as well as a graphene nano-transistor.

\section{\label{sec:tourwire} Tour Wire between Gold Electrodes}

In this section we will investigate the electrical properties of a phenylene
ethynylene oligomer, also popularly called a Tour wire. We will compare the
electrical properties of the molecule when it is symmetrically and
asymmetrically coupled with two Au(111) surfaces. In the symmetric system, as
illustrated in Fig.~\ref{fig:sym-device2}, the molecule is
connected with both gold surfaces through thiol bonds, whereas the asymmetric system
only has a thiol bond to one of them.

The system has previously been investigated experimentally by Kushmerick
\textit{et al.}\cite{Kushmerick-2002-PRL} and theoretically by Taylor
\textit{et al.}\cite{Taylor-2002-PRL}, and it has been found that the
asymmetrically coupled system shows strongly asymmetric
I--V characteristics\cite{Kushmerick-2002-PRL}.

The calculations by Taylor \textit{et al.} were based on
DFT-LDA, and the asymmetric behaviour could be related to the voltage drop in
the system. This system is therefore an excellent testing ground for our
semi-empirical model, since a correct description of the electrical properties requires not only a good model
for the zero-bias electronic structure, but also a good description of the bias-induced 
effects.

\subsection{Transmission Spectrum of the Symmetric Tour Wire Junction}
To setup the symmetric system we first relaxed the isolated Tour wire using
DFT-LDA\cite{ATK2008.10}. During the relaxation, passivating hydrogen atoms
were kept on the sulfur atoms. Afterwards, these hydrogen atoms were removed
and the two sulfur atoms placed at the FCC sites of two Au(111)-(3x3) surfaces.
The height of the S atom above the surface was 1.9 \AA\ (corresponding to an
Au--S distance of 2.53 \AA).

\begin{figure}[tbp]
\begin{center}
\includegraphics[width=\linewidth]{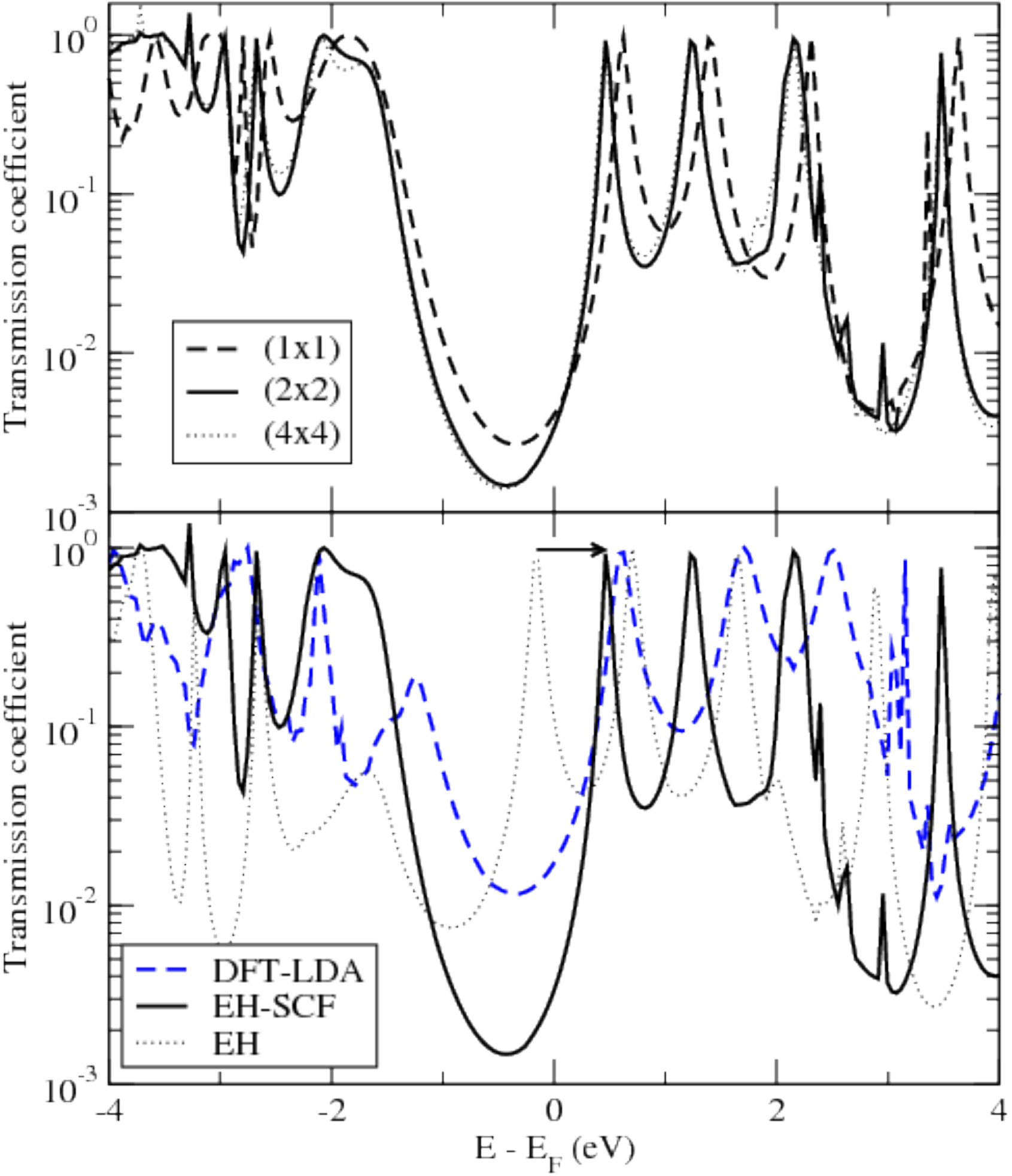}
\end{center}
\caption{(Color online) The upper plot shows the transmission spectrum of the
  symmetric Tour wire device, calculated with the EH-SCF model for three different k-point
  samplings. The lower plot shows transmission spectra calculated with a (2x2)
  k-point sampling using different models: EH-SCF (solid), EH without the
  Hartree term (dotted), and DFT-LDA (blue dashed). Energies are given relative
  to the Fermi level of the gold electrodes. }
\label{fig:tour_trans}
\end{figure}

We next set up the EH model with Hoffmann
parameters\cite{Hoffmann1978,Ammeter1978} and perform a self-consistent
calculation to obtain transmission spectra for different k-point sampling grids. The results are
shown in the upper plot of Fig.~\ref{fig:tour_trans}. In each case, the same k-point grid was
used for both the self-consistent and transmission calculation, and we see from the figure that 
using (1x1) k-point is insufficient while (2x2) and (4x4) k-points give almost
identical results. Thus, we will use a (2x2) k-point sampling grid for the remainder
of this study.

In the lower plot of Fig.~\ref{fig:tour_trans} we compare the transmission
spectra calculated with DFT-LDA, EH-SCF, and EH without the Hartree term of
Eq.~(\ref{eq:EH_hamiltonian}). For the DFT-LDA model we use similar parameters
as Taylor \textit{et al.}\cite{Taylor-2002-PRL}, except for the k-point sampling
which is (2x2) in the current study. The calculations in Ref.~\onlinecite{Taylor-2002-PRL} 
were performed with a (1x1) k-point
sampling, which is insufficient\cite{Thygesen-2005-PRB}, and thus the DFT-LDA
results in this study will differ from those by Taylor \textit{et al.}

For the EH calculation we see a peak in the transmission spectrum just around
the Fermi level of the gold electrodes. This peak arises from transmission
through the LUMO orbital of the Tour wire.

In the self-consistent EH calculation there will be a charge transfer from the gold
surface to the LUMO orbital, and we see that this
gives rise to a shift of the orbital by 1~eV, illustrated by the arrow in
Fig.~\ref{fig:tour_trans}.

For the DFT-LDA calculation we see that the LUMO peak is shifted further away
from the gold Fermi level, and the HOMO and LUMO peaks of the transmission
spectrum are placed almost symmetrically around the gold Fermi level. We also
note that the transmission coefficient at the Fermi level, corresponding to the
zero-bias conductance, is almost one order of magnitude higher within the
DFT-LDA model. We will discuss this further below.

We also note that Taylor \textit{et al.} find a LUMO level even
further away from the gold Fermi level; this is related to the insufficient
k-point sampling.

\subsection{I--V Characteristics of the Symmetric and Asymmetric Tour Wire Systems}

\begin{figure}[tbp]
\begin{center}
\includegraphics[width=\linewidth]{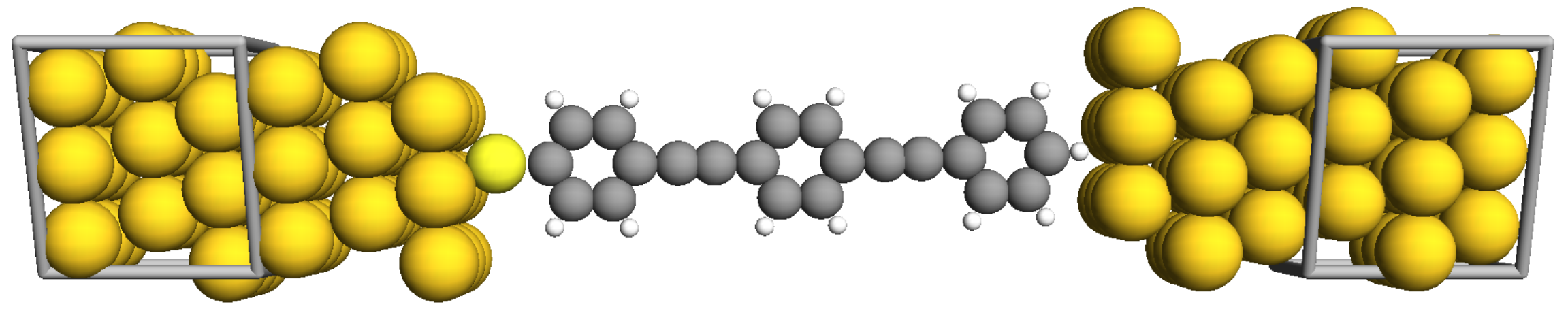}
\end{center}
\caption{(Color online) Geometry of the asymmetric system. The Tour wire is attached
  to the left gold electrode through a thiol bond, while the right end of the
  molecule is hydrogen-terminated and there is no chemical bond to the right gold
  electrode.  }
\label{fig:asym-device2}
\end{figure}

We will now study both the symmetric and asymmetric Tour wire system and compare their respective
I--V characteristics. The geometry of the asymmetric system is
illustrated in Fig.~\ref{fig:asym-device2}. The geometry is similar to that of
Fig.~\ref{fig:sym-device2}, except for the right-most sulfur atom which has been
replaced by a hydrogen atom with a C--H bond length of 1.1 \AA. The
distance between the hydrogen atom and the right gold surface is 1.5 \AA.

\begin{figure}[tbp]
\begin{center}
\includegraphics[width=\linewidth]{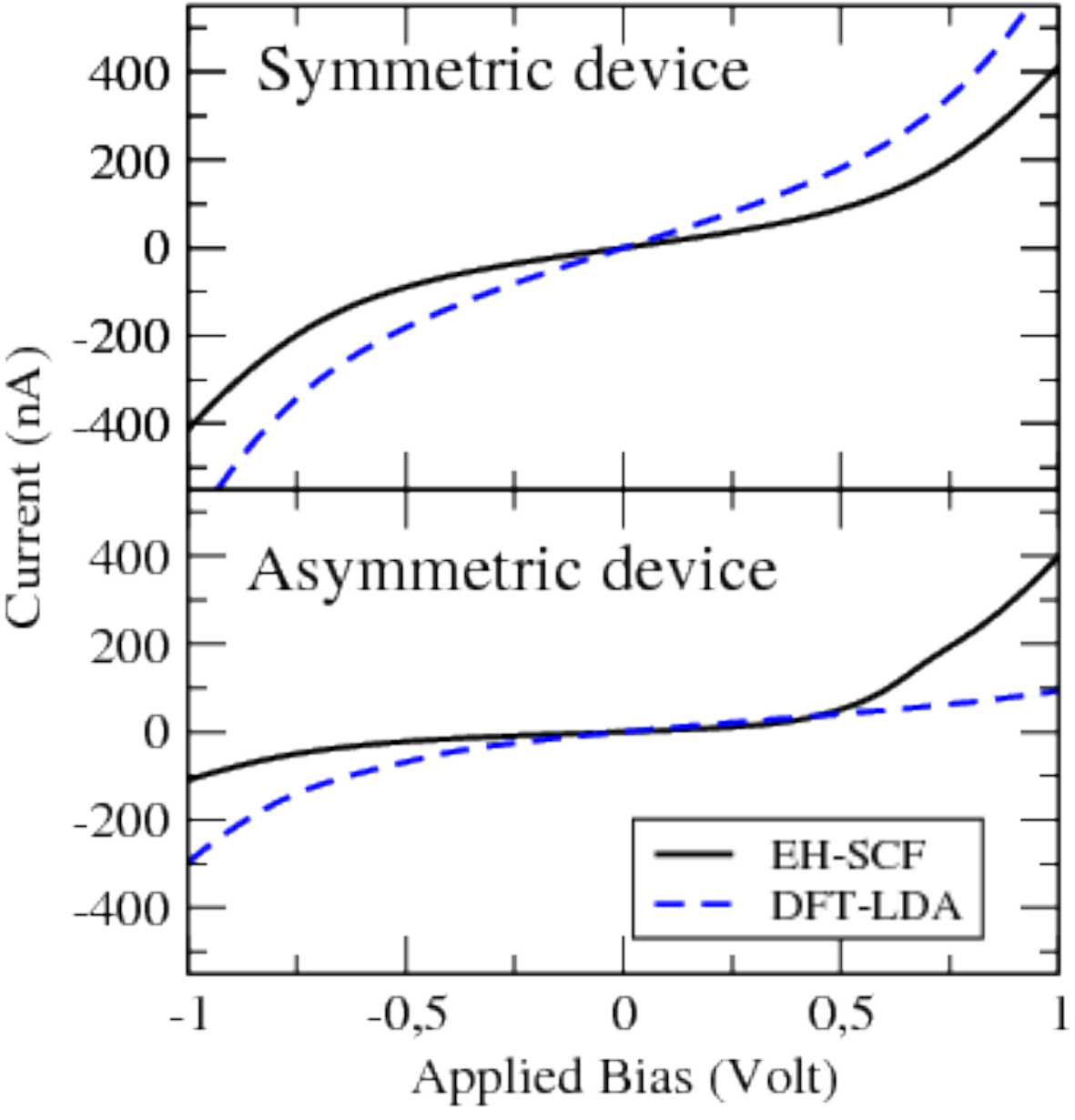}
\end{center}
\caption{(Color online) I--V characteristics of the symmetric (upper figure) and
  asymmetric (lower figure) Tour wire device. The positive current direction is from
  left to right. }
\label{fig:iv}
\end{figure}

We perform self-consistent calculations for both the symmetric and
asymmetric systems with the EH-SCF and DFT-LDA methods, and vary the bias
from --1 to +1~V in steps of 0.1~V. The results are shown in
Fig.~\ref{fig:iv}. For the symmetric device we obtain rather similar, symmetric
I--V characteristics for both the EH-SCF and DFT-LDA methods. The main difference is
that the zero-bias conductance is significantly higher with DFT-LDA,
reflecting the higher transmission coefficient at the Fermi level, as shown in
Fig.~\ref{fig:tour_trans}.

\begin{figure}[tbp]
\begin{center}
\includegraphics[width=\linewidth]{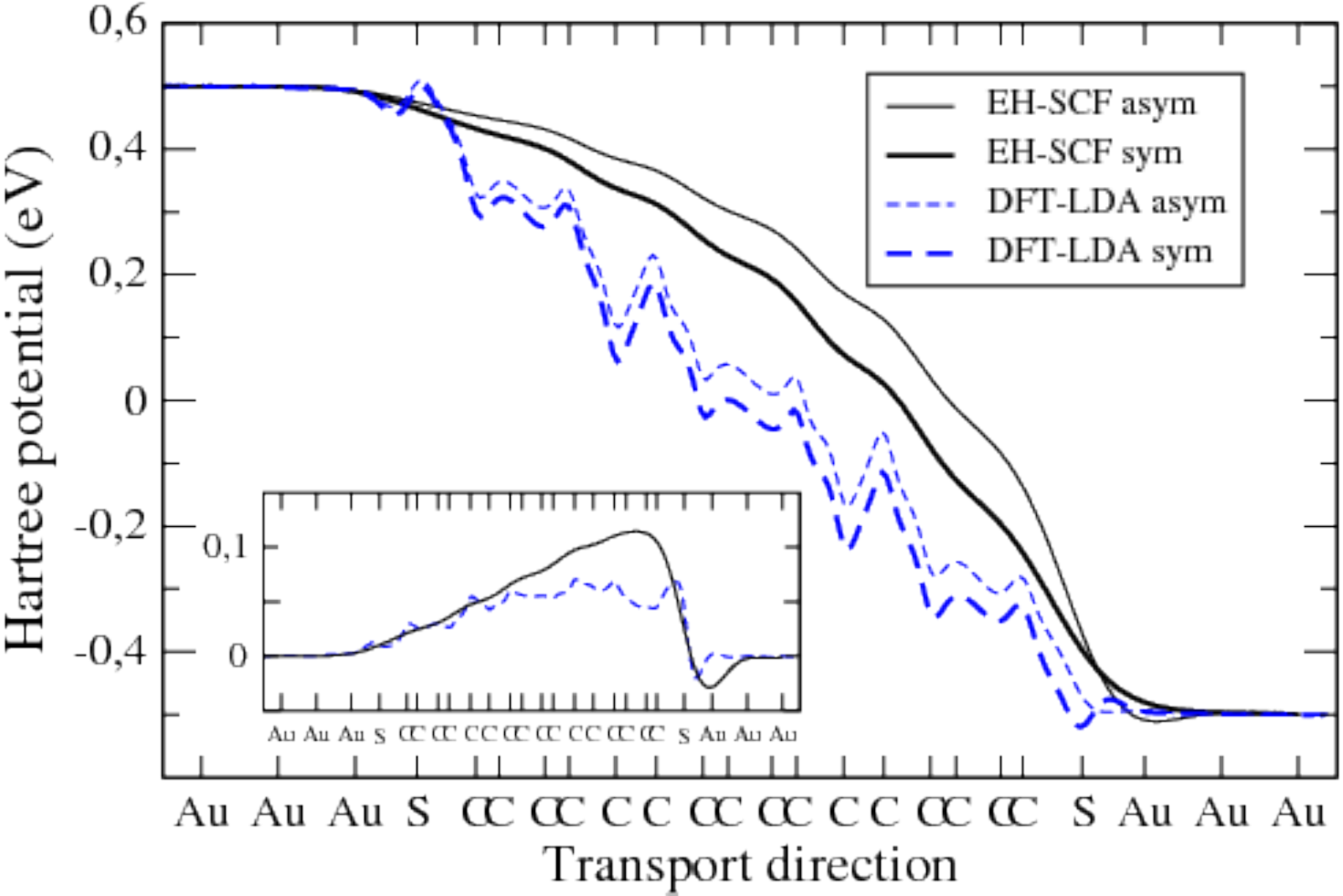}
\end{center}
\caption{(Color online) Voltage drop of the symmetric and asymmetric Tour wire systems
  along a line that goes through the two sulfur atoms in the symmetric system,
  for an applied bias of +1~V. The inset shows the voltage drop in the asymmetric
  system subtracted from the voltage drop in the symmetric system. Both plots show
  results calculated with the EH-SCF (solid) and DFT-LDA (dashed) models.}
\label{fig:voltage_drop}
\end{figure}

For the asymmetric device we see that both the DFT-LDA and EH-SCF models give
rise to rectification -- however, in opposite directions. Taylor
\textit{et al.}\cite{Taylor-2002-PRL} demonstrated that the rectification was
related to the voltage drop in the system, and we therefore in
Fig.~\ref{fig:voltage_drop} compare the voltage drops obtained with the two
methods. The EH-SCF voltage drop is smooth, since the charge density
is composed of a superposition of single rather broad Gaussians on
each atom. The DFT-LDA model shows atomic-scale details, however, as illustrated by the inset, the
relative change in the voltage drop between the asymmetric and symmetric system
is quite similar for the EH-SCF and DFT-LDA methods. Both methods reveal that in the asymmetric
system there is an additional voltage drop at the contact with the weak bond.
This is also one of the main results of  Taylor
\textit{et al.}\cite{Taylor-2002-PRL}.

The additional voltage drop at the weak contact means that the molecular levels
of the Tour wire mainly follow the electrochemical potential of the right
electrode\cite{Taylor-2002-PRL}. Since the voltage drop is similar for the
DFT-LDA and EH-SCF models, the difference in the I--V characteristics must be
related to the different electronic structure at zero bias in the two models.
Within the DFT-LDA model, the transport at the Fermi level is dominated by the
HOMO. At negative bias, the left electrode has a higher electrochemical
potential, and electrons from the occupied HOMO level can propagate to empty
states in the left electrode. Thus, for the DFT-LDA model, the current is highest
for a negative bias at the left electrode. For the EH-SCF model, on the
other hand, the transport at the Fermi level is dominated by the LUMO, and the
current in this case is highest for a positive bias at the left electrode.

Comparing with the experimental results of Kushmeric 
\textit{et al.}\cite{Kushmerick-2002-PRL}, we find that the EH-SCF rectification
direction agrees with the experimental rectification direction, while the
DFT-LDA model predicts rectification in the opposite direction. 
We note that the rectification
direction obtained with our DFT-LDA model is similar to the results of Taylor
\textit{et al.}\cite{Taylor-2002-PRL}.

Thus, for this system it seems that the EH-SCF model is in better agreement
with the experimental results, compared to the DFT-LDA model. The
example shows that the EH-SCF model gives a very good description
of both the electronic structure and the
voltage drop in the system. The comparisons
between the two methods also illustrates how small variations in the positions of the HOMO
and LUMO levels may change the electrical properties of the Tour wire device.

\section{\label{sec:graphene} Z-Shaped Graphene Nano-Transistor}

\begin{figure}[tbp]
\centering
\includegraphics[width=.91\linewidth]{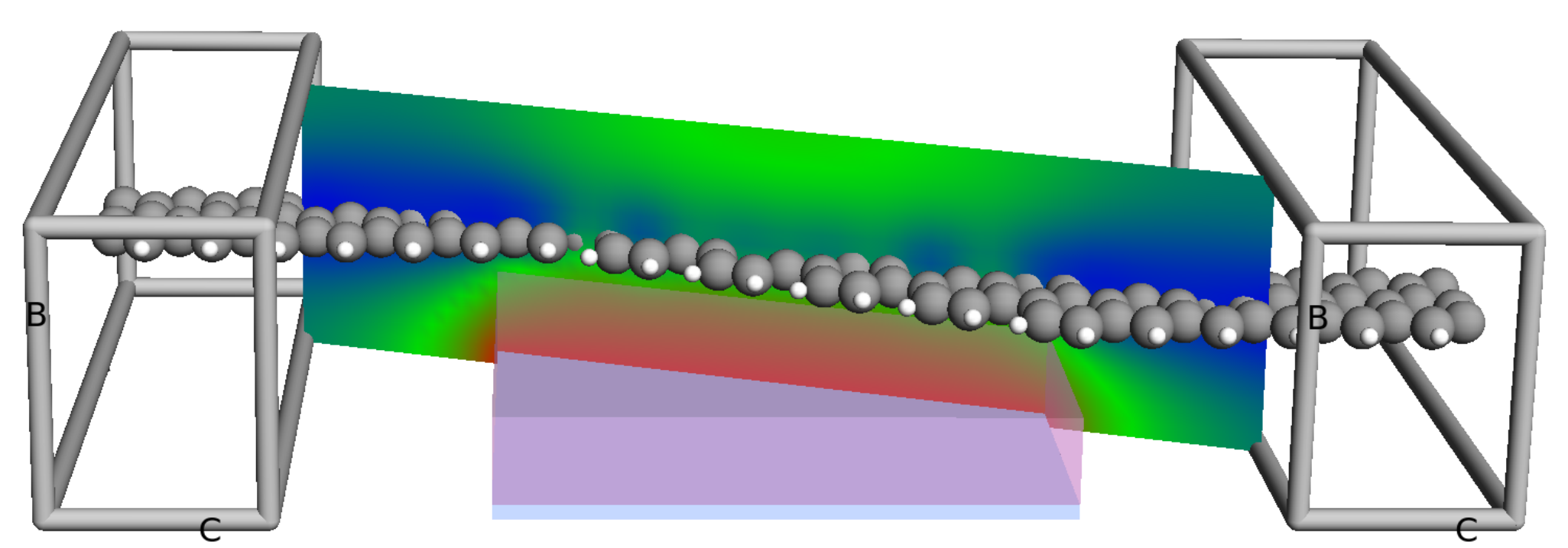}
\includegraphics[width=.07\linewidth]{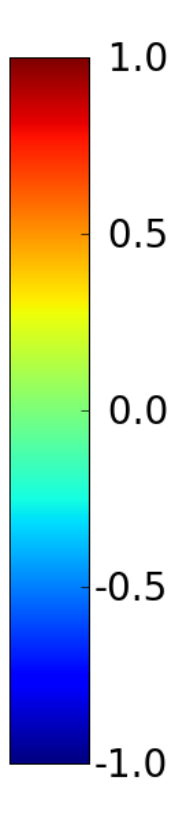}
\caption{(Color online) Graphene nano-transistor consisting of two metallic zigzag
  nanoribbons connected by a semiconducting armchair 
  ribbon. The nanoribbons are passivated with hydrogen, and the width of the
  ribbons are is 7~\AA. The device
  is sitting on top of a dielectric and the transport is controlled by an
  electrostatic back-gate. The contour plot illustrates the Hartree
  potential for a gate potential of --1~V. }
\label{fig:graphene}
\end{figure}

In this section we will compare the electrical properties of a short (34~\AA)
and long (86~\AA) graphene nano-transistor. The system consists of two electrodes consisting of metallic,
zigzag-edge graphene nanoribbons connected through a semiconducting armchair-edge
central ribbon. The system is placed 1.4~\AA\ above a dielectric material
with dielectric constant $\epsilon = 4 \epsilon_0$, corresponding to SiO$_2$. The
dielectric is 3~\AA\ thick, and below the dielectric there is an electrostatic
gate. The geometry of the short system is illustrated in
Fig.~\ref{fig:graphene}. A similar system was investigated by Yan
\textit{et al.}\cite{Yan-2007-NL} using DFT-LDA.

For the calculation we use EH parameters from Ref.~\onlinecite{KienleI2006} which
were derived by fitting to a reference band structure of a graphene sheet calculated
with DFT-LDA.  With these parameters, we find a band gap of the central ribbon of
2.2 eV, in agreement with DFT-LDA calculations, which illustrates the transferability
of the EH parameters from 2D graphene to a 1D graphene nanoribbon.

\subsection{Transmission Spectrum}

\begin{figure}[tbp]
\begin{center}
\includegraphics[width=\linewidth]{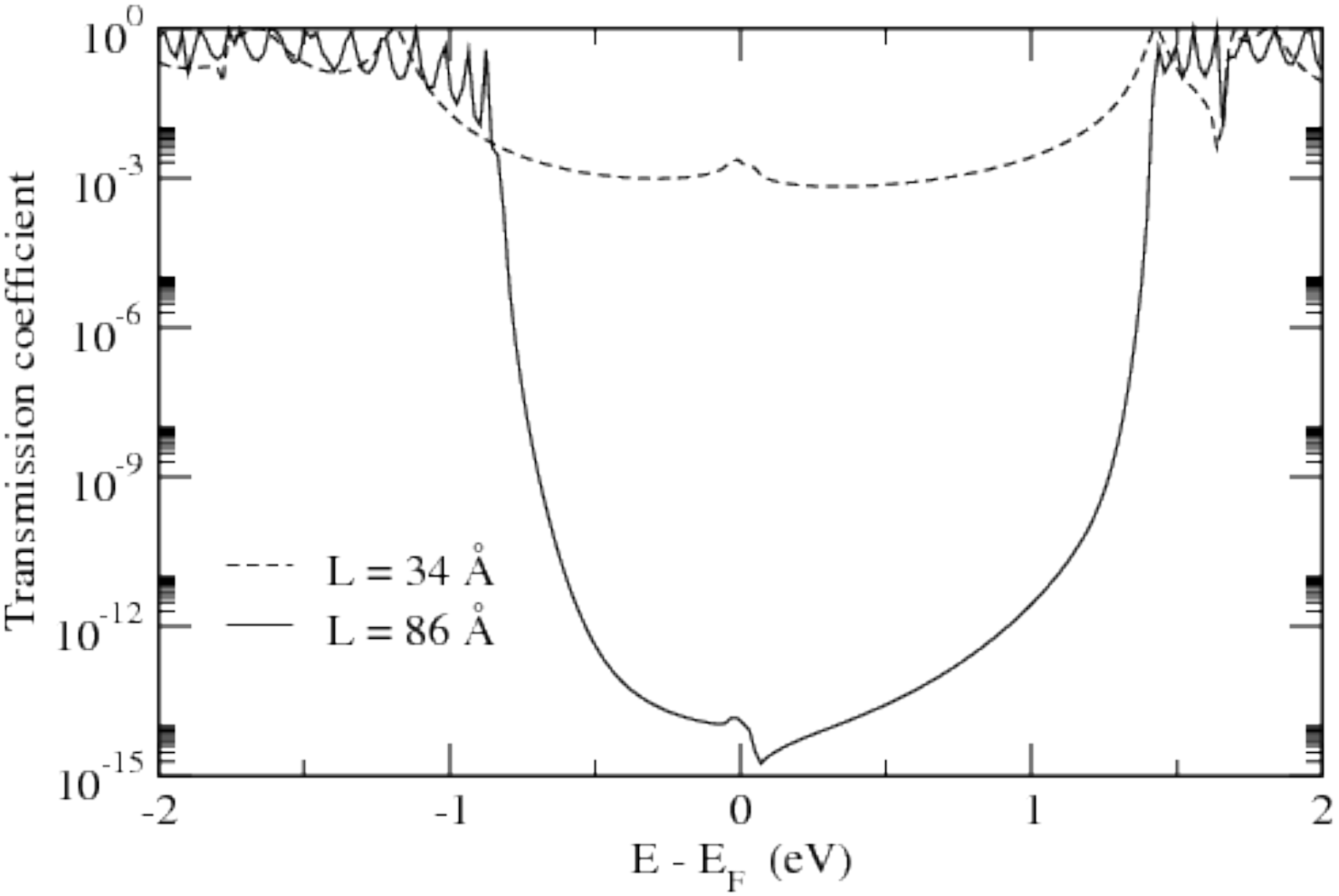}
\end{center}
\caption{Zero-bias transmission spectrum for the short (dashed) and long (solid) graphene
  device. Energies are relative to the Fermi level of the electrodes.}
\label{fig:transgraphene}
\end{figure}

Fig.~\ref{fig:transgraphene} shows the transmission spectrum for both the long
and the short system when there is no applied bias and zero gate potential. The shape of the transmission spectrum is directly related to the electronic structure of
the central semiconducting ribbon. 

The transmission is strongly reduced in the energy region from --0.7 to 1.5~eV,
corresponding to the band gap of the central armchair ribbon. Since there
are no energy levels in this interval, the electrons
must tunnel in order to propagate across the junction. For the longer device the
electrons must tunnel a longer distance, and thus the transmission is more
strongly reduced.

Outside the band gap, the transmission is close to 1 and shows a number of
oscillations. Since the central ribbon has a finite
length, it resembles a molecule with a number of discrete energy levels. The
levels give rise to peaks in the transmission spectrum, and since the longer
system has more energy levels, the peaks are more closely spaced there.

In the following section we will see how this difference in the transmission spectrum
gives rise to qualitatively different transport mechanisms in the two 
devices.

\subsection{Transistor Characteristics}

\begin{figure}[tbp]
\begin{center}
\includegraphics[width=\linewidth]{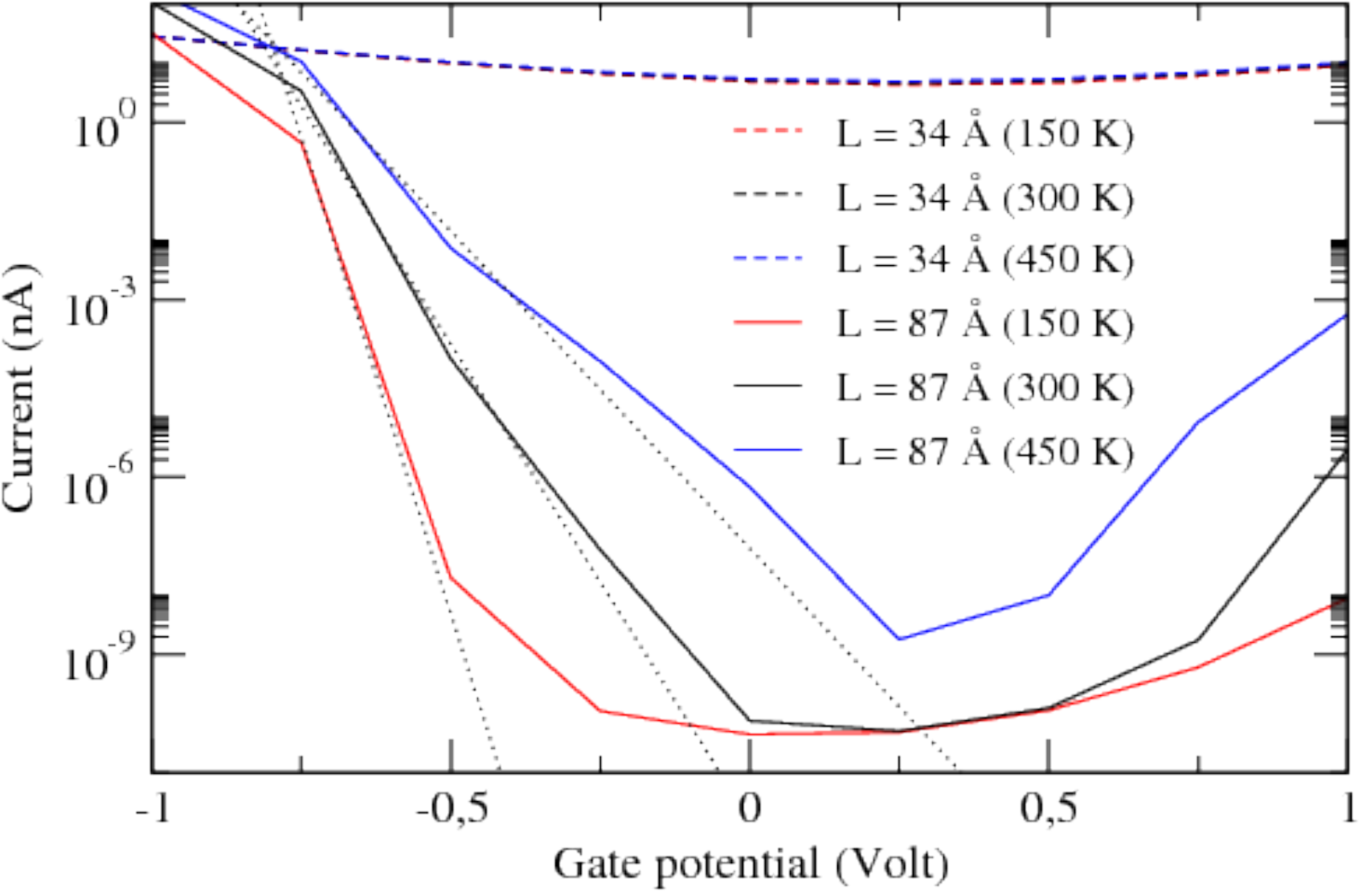}
\end{center}
\caption{(Color online) The tunneling current for a source--drain voltage of 0.2~V, as a
  function of the gate potential for the long (solid) and short (dashed)
  graphene device, respectively. Three different values of the electron temperature in 
  the electrodes were considered: 150~K, 300~K and 450~K. The dotted lines illustrate the $1/k_\mathrm{B} T$ slope
  for the different temperatures.}
\label{fig:gateiv_t}
\end{figure}

We now calculate the current for an applied source--drain voltage of 0.2~V
as a function of the applied gate potential. Fig.~\ref{fig:gateiv_t} shows the
current for the long and short devices, respectively, for gate potentials in the range --1 to 1~V,
for different electrode temperatures. We see that
for the short device there is only a small effect of the gate potential and
electron temperature, while for the long device the conductance falls off
exponentially, reaching a minimum in the range 0 to 0.5~V. Moreover, the current is
strongly temperature-dependent.

The lack of temperature-dependence for the short device shows that the transport is
completely dominated by electron tunneling. For the long device, on the other hand, there is a
strong temperature dependence, and in this case the electron transport is
dominated by thermionic emission. The dotted lines illustrate the $1/k_\mathrm{B} T$
slope expected for thermionic emission. We see that in the gate voltage range from
--0.25 to --0.75~V, the I--V characterics follow these slopes well.

Fig.~\ref{fig:graphene} also shows the electrostatic profile through the
device. We see that the gate potential is almost perfectly screened by the
graphene ribbon, i.e.\ the gate potential does not penetrate through the central
ribbon. This means that for a layered structure, only the first layer
would be strongly affected by the back-gate. This has some implications also for gated nanotube
devices. In such a device, only the atoms facing the gate
electrode will be strongly influenced, and this explains why in
Ref.~\onlinecite{sorensen2009} we found that the transport in the device was dominated by
tunneling even though the nanotube was 110~\AA\ long, and thus longer than the graphene
junctions studied in this paper. Thus, to obtain efficient gating of a nanotube,
the gate electrode must wrap around the tube.

\section{\label{sec:summary} Conclusions}

In this paper we have introduced a new semi-empirical model for electron transport in nano-devices.
The model is based on the Extended H{\"u}ckel method that extends the work
by Zahid \textit{et al.}\cite{Zahid2005} to give a more complete description of the
electrostatic interactions in the device. In particular, the position of the
electrode Fermi level and the charge transfer between the contacts and the device are
calculated self-consistently.

Compared to DFT-based transport methods, the main advantage of our new method is that it is
computationally less expensive, as well as having the option of adjusting parameters to
reproduce experimental data or computationally very demanding many-body electronic
structure methods.

The model includes a self-consistent Hartree potential which takes into account
the effect of an external bias as well as continuum dielectric
regions and external electrostatic gates.

We used the model to study a Tour wire between gold electrodes, and found that
the voltage drop in the device compares well with \textit{ab initio} results, while
the calculated current--voltage characteristics qualitatively agree better with experimental findings
than the corresponding DFT-LDA results do.

We also considered a graphene nano-transistor, and our study illustrated how the 
transport mechanism changes from tunnelling to thermionic emission as the device is made longer.

These applications show that the new method can give an accurate description of a
broad range of nano-scale devices. With its favorable computational speed, it is
a good complement to \textit{ab initio}-based transport
methods.
 
\begin{acknowledgments}\label{sec:acknowledgements}

This work was supported by the Danish Council for Strategic Research 'NABIIT'
under Grant No.\ 2106-04-0017, ``Parallel Algorithms for Computational
Nano-Science'', and European Commission STREP project No.\ MODECOM
``NMP-CT-2006-016434'', EU.

\end{acknowledgments}

%############################################################

\bibliography{paper}    % Bibliography

\end{document}